%% file: main.tex
\pdfoutput=1
\documentclass[12pt,a4paper]{article}
\usepackage{ifthen} % for conditional statements
\newboolean{pdflatex}
\setboolean{pdflatex}{true} % False for eps figures

\newboolean{articletitles}
\setboolean{articletitles}{true} % False removes titles in references

\newboolean{uprightparticles}
\setboolean{uprightparticles}{false} %True for upright particle symbols

\newboolean{inbibliography}
\setboolean{inbibliography}{false} %True once you enter the bibliography

\usepackage{listings} %for verbatim with line wrap (PRL justification)
\input{PRL-defs}

\input{preamble}
\usepackage{rotating}
\usepackage{pdflscape}
\usepackage[section]{placeins}
\begin{document}

%%%%%%%%%%%%%%%%%%%%%%%%%
%%%%% Title     %%%%%%%%%
%%%%%%%%%%%%%%%%%%%%%%%%%
\renewcommand{\thefootnote}{\fnsymbol{footnote}}
\setcounter{footnote}{1}
\input{title-LHCb-PAPER}

\renewcommand{\thefootnote}{\arabic{footnote}}
\setcounter{footnote}{0}

%%%%%%%%%%%%%%%%%%%%%%%%%
%%%%% Main text %%%%%%%%%
%%%%%%%%%%%%%%%%%%%%%%%%%

\pagestyle{plain} % restore page numbers for the main text
\setcounter{page}{1}
\pagenumbering{arabic}

%% Uncomment during review phase.
%% Comment before a final submission.
%\linenumbers

\input{Introduction/introduction}

\input{Models/models}

\input{Results/results}

\input{Conclusions/conclusions}

\input{acknowledgements}

\addcontentsline{toc}{section}{References}
\setboolean{inbibliography}{true}
\bibliographystyle{LHCb}
\bibliography{main,LHCb-PAPER,LHCb-CONF,LHCb-DP,LHCb-TDR}

\clearpage

\newpage

\input{LHCb_Authorship_12-Mar-2019.tex}

\end{document}

%% file: PRL-defs.tex
\def\paperauthors{LHCb collaboration} % Leave as is for PAPER and CONF
\def\paperasciititle{Observation of several sources of CP violation in B+ -> pi_pi+pi- decays} % Set ASCII title here
\def\papertitle{Observation of several sources of \CP violation in \decay{\Bp}{\pip\pip\pim} decays} % Latex formatted title
\def\lhcbpapernumber{LHCb-PAPER-2019-018}
\def\cernpreprintnumber{CERN-EP-2019-156}
\def\paperkeywords{{High Energy Physics}, {LHCb}} % Comma separated list
\def\papercopyright{\the\year\ CERN for the benefit of the LHCb collaboration} % new since 9/Apr/2018
\def\paperlicence{CC-BY-4.0 licence}
\def\paperlicenceurl{https://creativecommons.org/licenses/by/4.0/}

%% file: preamble.tex
\usepackage[top=1in, bottom=1.25in, left=1in, right=1in]{geometry}

\usepackage{microtype}
\usepackage{lineno}  % for line numbering during review
\usepackage{xspace} % To avoid problems with missing or double spaces after
\usepackage{caption} %these three command get the figure and table captions automatically small

\usepackage{graphicx}  % to include figures (can also use other packages)
\usepackage{color}
\usepackage{colortbl}
\graphicspath{{./figs/}} % Make Latex search fig subdir for figures

\usepackage{amsmath} % Adds a large collection of math symbols
\usepackage{amssymb}
\usepackage{amsfonts}
\usepackage{upgreek} % Adds in support for greek letters in roman typeset

\newcommand*\patchAmsMathEnvironmentForLineno[1]{%
\expandafter\let\csname old#1\expandafter\endcsname\csname #1\endcsname
\expandafter\let\csname oldend#1\expandafter\endcsname\csname
end#1\endcsname
 \renewenvironment{#1}%
   {\linenomath\csname old#1\endcsname}%
   {\csname oldend#1\endcsname\endlinenomath}%
}
\newcommand*\patchBothAmsMathEnvironmentsForLineno[1]{%
  \patchAmsMathEnvironmentForLineno{#1}%
  \patchAmsMathEnvironmentForLineno{#1*}%
}
\AtBeginDocument{%
\patchBothAmsMathEnvironmentsForLineno{equation}%
\patchBothAmsMathEnvironmentsForLineno{align}%
\patchBothAmsMathEnvironmentsForLineno{flalign}%
\patchBothAmsMathEnvironmentsForLineno{alignat}%
\patchBothAmsMathEnvironmentsForLineno{gather}%
\patchBothAmsMathEnvironmentsForLineno{multline}%
\patchBothAmsMathEnvironmentsForLineno{eqnarray}%
}

\usepackage{hyperxmp}

\usepackage[pdftex,
            pdfauthor={\paperauthors},
            pdftitle={\paperasciititle},
            pdfkeywords={\paperkeywords},
            pdfcopyright={Copyright (C) \papercopyright},
            pdflicenseurl={\paperlicenceurl}]{hyperref}

\usepackage[all]{hypcap} % Internal hyperlinks to floats.

\input{lhcb-symbols-def} % Add in the predefined LHCb symbols

\usepackage{cite} % Allows for ranges in citations
\usepackage{mciteplus}

%% file: lhcb-symbols-def.tex
\usepackage{xspace} 
\usepackage{upgreek}

\def\MagUp {\mbox{\em Mag\kern -0.05em Up}\xspace}

\ifthenelse{\boolean{uprightparticles}}%
{

 \def\Ppi         {\ensuremath{\uppi}\xspace}

 \def\PDelta      {\ensuremath{\Delta}\xspace}                 
 \def\PXi      {\ensuremath{\Xi}\xspace}                 
 \def\PLambda      {\ensuremath{\Lambda}\xspace}                 
 \def\PSigma      {\ensuremath{\Sigma}\xspace}                 
 \def\POmega      {\ensuremath{\Omega}\xspace}                 
 \def\PUpsilon      {\ensuremath{\Upsilon}\xspace}                 
 
 \def\PB      {\ensuremath{\mathrm{B}}\xspace}                 
                  
 \def\PD      {\ensuremath{\mathrm{D}}\xspace}

 \def\PK      {\ensuremath{\mathrm{K}}\xspace}

 \def\Pb      {\ensuremath{\mathrm{b}}\xspace}

 \def\Pi      {\ensuremath{\mathrm{i}}\xspace}

 \def\Pp      {\ensuremath{\mathrm{p}}\xspace}

}
{

 \def\Ppi         {\ensuremath{\pi}\xspace}

 \mathchardef\PDelta="7101
 \mathchardef\PXi="7104
 \mathchardef\PLambda="7103
 \mathchardef\PSigma="7106
 \mathchardef\POmega="710A
 \mathchardef\PUpsilon="7107
                  
 \def\PB      {\ensuremath{B}\xspace}                 
                  
 \def\PD      {\ensuremath{D}\xspace}

 \def\PK      {\ensuremath{K}\xspace}

 \def\Pb      {\ensuremath{b}\xspace}

 \def\Pi      {\ensuremath{i}\xspace}

 \def\Pp      {\ensuremath{p}\xspace}

}

\makeatletter
\ifcase \@ptsize \relax% 10pt
  \newcommand{\miniscule}{\@setfontsize\miniscule{4}{5}}% \tiny: 5/6
\or% 11pt
  \newcommand{\miniscule}{\@setfontsize\miniscule{5}{6}}% \tiny: 6/7
\or% 12pt
  \newcommand{\miniscule}{\@setfontsize\miniscule{5}{6}}% \tiny: 6/7
\fi
\makeatother

\DeclareRobustCommand{\optbar}[1]{\shortstack{{\miniscule (\rule[.5ex]{1.25em}{.18mm})}
  \\ [-.7ex] $#1$}}

\def\bquark    {{\ensuremath{\Pb}}\xspace}

\def\pion   {{\ensuremath{\Ppi}}\xspace}

\def\pip    {{\ensuremath{\pion^+}}\xspace}
\def\pim    {{\ensuremath{\pion^-}}\xspace}
\def\pipm   {{\ensuremath{\pion^\pm}}\xspace}

\def\kaon    {{\ensuremath{\PK}}\xspace}
  \def\Kbar    {{\kern 0.2em\overline{\kern -0.2em \PK}{}}\xspace}

\def\KorKbar    {\kern 0.18em\optbar{\kern -0.18em K}{}\xspace}

\def\Kp      {{\ensuremath{\kaon^+}}\xspace}

\def\Dbar    {{\kern 0.2em\overline{\kern -0.2em \PD}{}}\xspace}

\def\DorDbar    {\kern 0.18em\optbar{\kern -0.18em D}{}\xspace}

\def\Dzb     {{\ensuremath{\Dbar{}^0}}\xspace}

\def\B       {{\ensuremath{\PB}}\xspace}
\def\Bbar    {{\ensuremath{\kern 0.18em\overline{\kern -0.18em \PB}{}}}\xspace}

\def\BorBbar    {\kern 0.18em\optbar{\kern -0.18em B}{}\xspace}
\def\Bz      {{\ensuremath{\B^0}}\xspace}

\def\Bu      {{\ensuremath{\B^+}}\xspace}
\def\Bub     {{\ensuremath{\B^-}}\xspace}
\def\Bp      {{\ensuremath{\Bu}}\xspace}
\def\Bm      {{\ensuremath{\Bub}}\xspace}

  \def\Y#1S{\ensuremath{\PUpsilon{(#1S)}}\xspace}% no space before {...}!

\def\proton      {{\ensuremath{\Pp}}\xspace}

\def\Lbar        {{\ensuremath{\kern 0.1em\overline{\kern -0.1em\PLambda}}}\xspace}
\def\LorLbar    {\kern 0.18em\optbar{\kern -0.18em \PLambda}{}\xspace}

\newcommand{\decay}[2]{\ensuremath{#1\!\to #2}\xspace}         % {\Pa}{\Pb \Pc}

\def\to                 {\ensuremath{\rightarrow}\xspace}

\def\CP                {{\ensuremath{C\!P}}\xspace}

\def\AT#1     {\ensuremath{A_{\mathrm{T}}^{#1}}\xspace}           % 2

\def\C#1      {\ensuremath{\mathcal{C}_{#1}}\xspace}                       % 9
\def\Cp#1     {\ensuremath{\mathcal{C}_{#1}^{'}}\xspace}                    % 7
\def\Ceff#1   {\ensuremath{\mathcal{C}_{#1}^{\mathrm{(eff)}}}\xspace}        % 9  
\def\Cpeff#1  {\ensuremath{\mathcal{C}_{#1}^{'\mathrm{(eff)}}}\xspace}       % 7
\def\Ope#1    {\ensuremath{\mathcal{O}_{#1}}\xspace}                       % 2
\def\Opep#1   {\ensuremath{\mathcal{O}_{#1}^{'}}\xspace}                    % 7

\newcommand{\tev}{\ifthenelse{\boolean{inbibliography}}{\ensuremath{~T\kern -0.05em eV}}{\ensuremath{\mathrm{\,Te\kern -0.1em V}}}\xspace}
\newcommand{\gev}{\ensuremath{\mathrm{\,Ge\kern -0.1em V}}\xspace}
\newcommand{\mev}{\ensuremath{\mathrm{\,Me\kern -0.1em V}}\xspace}
\newcommand{\kev}{\ensuremath{\mathrm{\,ke\kern -0.1em V}}\xspace}
\newcommand{\ev}{\ensuremath{\mathrm{\,e\kern -0.1em V}}\xspace}
\newcommand{\gevc}{\ensuremath{{\mathrm{\,Ge\kern -0.1em V\!/}c}}\xspace}
\newcommand{\mevc}{\ensuremath{{\mathrm{\,Me\kern -0.1em V\!/}c}}\xspace}
\newcommand{\gevcc}{\ensuremath{{\mathrm{\,Ge\kern -0.1em V\!/}c^2}}\xspace}
\newcommand{\gevgevcccc}{\ensuremath{{\mathrm{\,Ge\kern -0.1em V^2\!/}c^4}}\xspace}
\newcommand{\mevcc}{\ensuremath{{\mathrm{\,Me\kern -0.1em V\!/}c^2}}\xspace}

\def\invfb   {\ensuremath{\mbox{\,fb}^{-1}}\xspace}

\def\gsim{{~\raise.15em\hbox{$>$}\kern-.85em
          \lower.35em\hbox{$\sim$}~}\xspace}
\def\lsim{{~\raise.15em\hbox{$<$}\kern-.85em
          \lower.35em\hbox{$\sim$}~}\xspace}

\def\tell1  {TELL1\xspace}
\def\ukl1   {UKL1\xspace}

\newcommand{\eg}{\mbox{\itshape e.g.}\xspace}

%% file: title-LHCb-PAPER.tex
% $Id: title-LHCb-PAPER.tex 118821 2018-04-09 19:15:15Z pkoppenb $
% ===============================================================================
% Purpose: LHCb-PAPER journal paper title page template
% Author:
% Created on: 2010-09-25
% ===============================================================================

%%%%%%%%%%%%%%%%%%%%%%%%%
%%%%%  TITLE PAGE  %%%%%%
%%%%%%%%%%%%%%%%%%%%%%%%%
\begin{titlepage}
\pagenumbering{roman}

% Header ---------------------------------------------------
\vspace*{-1.5cm}
\centerline{\large EUROPEAN ORGANIZATION FOR NUCLEAR RESEARCH (CERN)}
\vspace*{1.5cm}
\noindent
\begin{tabular*}{\linewidth}{lc@{\extracolsep{\fill}}r@{\extracolsep{0pt}}}
\ifthenelse{\boolean{pdflatex}}% Logo format choice
{\vspace*{-1.5cm}\mbox{\!\!\!\includegraphics[width=.14\textwidth]{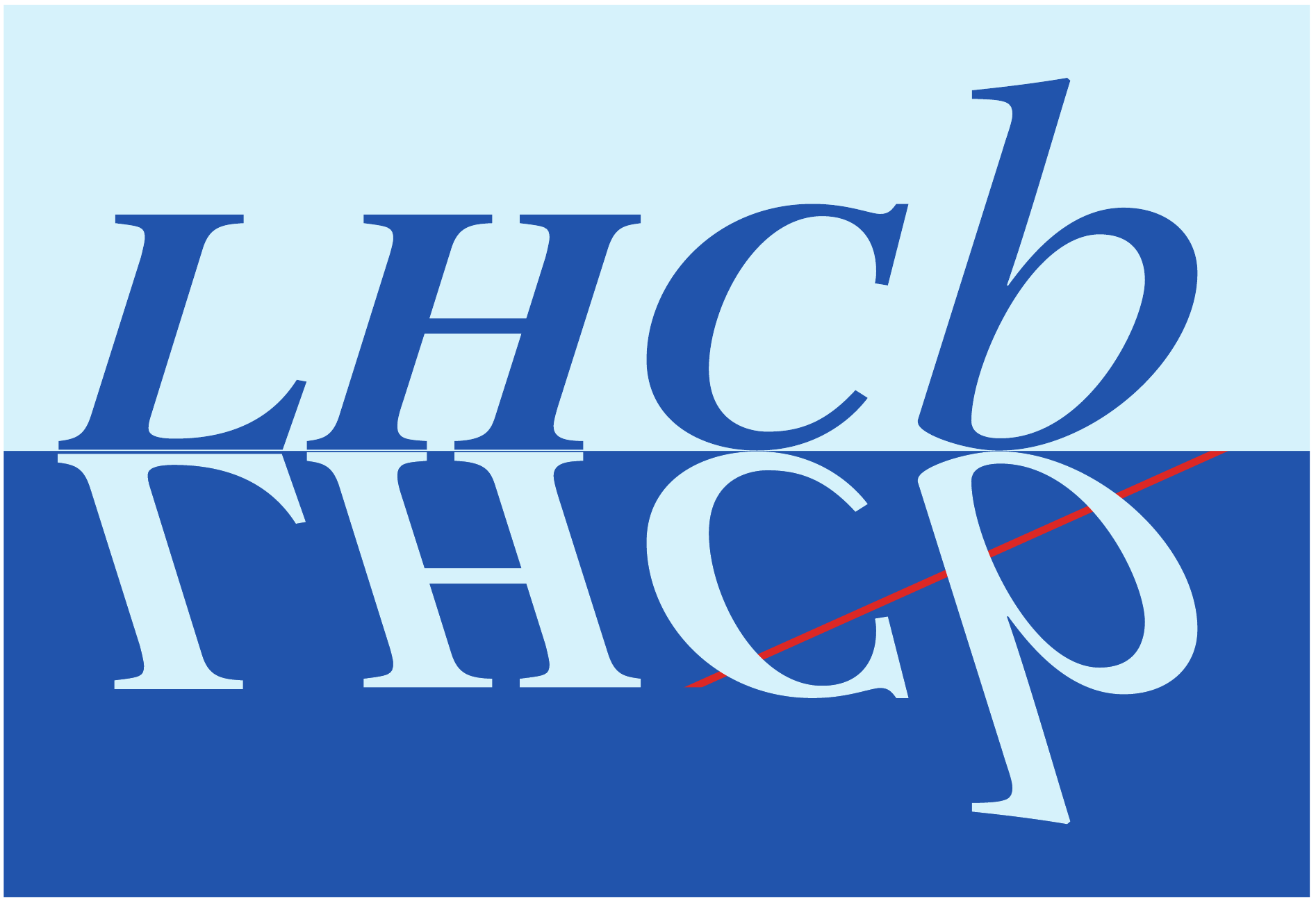}} & &}%
{\vspace*{-1.2cm}\mbox{\!\!\!\includegraphics[width=.12\textwidth]{lhcb-logo.eps}} & &}%
\\
 & & \cernpreprintnumber \\  % ID
 & & \lhcbpapernumber \\  % ID
% & & \today \\ % Date - Can also hardwire e.g.: 23 March 2010
% & & 12 September 2019 \\ % Date - Can also hardwire e.g.: 23 March 2010
 & & 23 January 2020 \\ % Date - Can also hardwire e.g.: 23 March 2010
 & & \\
% not in paper \hline
\end{tabular*}

\vspace*{4.0cm}

% Title --------------------------------------------------
{\normalfont\bfseries\boldmath\huge
\begin{center}
% DO NOT EDIT HERE. Instead edit macro in main.tex to keep metadata correct
  \papertitle
\end{center}
}

\vspace*{2.0cm}

% Authors -------------------------------------------------
\begin{center}
%In the footnote, replace 'paper' by 'Letter' in case of submission to PRL or PLB
% Edit macro in main.tex to keep metadata correct
%\paperauthors\footnote{Authors are listed at the end of this paper.}
\paperauthors\footnote{Authors are listed at the end of this Letter.}
\end{center}

\vspace{\fill}

% Abstract -----------------------------------------------
\input{abstract}

\vspace*{2.0cm}

\begin{center}
  %Submitted to Phys.~Rev.~Lett
  Published in Phys. Rev. Lett. 124 (2020) 031801
\end{center}

\vspace{\fill}

{\footnotesize
% Edit macro in main.tex to keep metadata correct
\centerline{\copyright~\papercopyright. \href{\paperlicenceurl}{\paperlicence}.}}
\vspace*{2mm}

\end{titlepage}

%%%%%%%%%%%%%%%%%%%%%%%%%%%%%%%%
%%%%%  EOD OF TITLE PAGE  %%%%%%
%%%%%%%%%%%%%%%%%%%%%%%%%%%%%%%%

%  empty page follows the title page ----
\newpage
\setcounter{page}{2}
\mbox{~}
%\newpage
%
%% Author List ----------------------------
%%  You need to get a new author list!
%\input{LHCb_authorlist.tex}
%
%The author list for journal publications is provided by the Membership Committee shortly after 'approval to go to paper' has been given.
%%It will be made available on the page
%%\verb!http://www.physik.uzh.ch/~strauman/forMemCo/LHCb-PAPER-XXXX-XXX/! .
%It will be sent to you by email shortly after a paper number has beens assigned.
%The author list should be included already at first circulation,
%to allow new members of the collaboration to verify whether they have been included correctly.
%Occasionally a misspelled name is corrected or associated institutions become full members.
%In that case, a new author list will be sent to you.
%In case line numbering doesn't work well after including the authorlist, try moving the \verb!\bigskip! after the last author to a separate line.
%
%
%The authorship for Conference Reports should be ``The LHCb
%  collaboration'', with a footnote giving the name(s) of the contact
%  author(s), but without the full list of collaboration names.

\cleardoublepage

%% file: abstract.tex
\begin{abstract}
  \noindent
  Observations are reported of different sources of \CP violation from an amplitude analysis of \decay{\Bp}{\pip\pip\pim} decays, based on a data sample corresponding to an integrated luminosity of $3 \invfb$ of $\proton\proton$ collisions recorded with the LHCb detector.
  A large \CP asymmetry is observed in the decay amplitude involving the tensor $f_2(1270)$ resonance, and
  in addition significant \CP violation is found in the $\pip\pim$ S-wave at low invariant mass.
  The presence of \CP violation related to interference between the $\pip\pim$ S-wave and the P-wave \decay{\Bp}{\rho(770)^0\pip} amplitude is also established; this causes large local asymmetries but cancels when integrated over the phase space of the decay.
  The results provide both qualitative and quantitative new insights into \CP-violation effects in hadronic \B decays.
\end{abstract}

%% file: Introduction/introduction.tex
Violation of symmetry under the combined charge-conjugation and parity-transformation operations, \CP violation, gives rise to differences between matter and antimatter.
Violation of \CP symmetry can occur in the amplitudes that describe hadron decay, in neutral hadron mixing, or in the interference between mixing and decay (for a review, see, \eg, Ref.~\cite{Gershon:2016fda}).
For charged mesons,
only \CP violation in decay is possible, where an asymmetry in particle and antiparticle decay rates can arise when two or more different amplitudes contribute to a transition.
In particular, the phase of each complex amplitude can be decomposed into a weak phase, which changes sign under \CP, and a strong phase, which is \CP invariant.
Differences in both the weak and strong phases of the contributing amplitudes are required for an asymmetry to occur.

In the Standard Model (SM), weak phases arise from the elements of the Cabibbo--Kobayashi--Maskawa matrix~\cite{Cabibbo:1963yz,Kobayashi:1973fv} that are associated with quark-level transition amplitudes.
Decays of \B hadrons that do not contain any charm quarks in the final state, such as \decay{\Bp}{\pip\pip\pim}, are of particular interest as both tree-level and loop-level amplitudes are expected to contribute with comparable magnitudes, so that large \CP-violation effects are possible.
Indeed, significant asymmetries have been observed in the two-body \decay{\Bz}{\Kp\pim}~\cite{Lees:2012mma,Duh:2012ie,LHCb-PAPER-2018-006} and \decay{\Bz}{\pip\pim}~\cite{Lees:2012mma,Adachi:2013mae,LHCb-PAPER-2018-006} decays.
In two-body decays, nontrivial strong phases can arise from rescattering or other hadronic effects.
In three-body or multibody decays, variation of the strong phase is also expected due to the intermediate resonance structure, and hence amplitude analyses can provide additional sensitivity to \CP-violation effects.

Analysis of the distribution of \decay{\Bp}{\pip\pip\pim} decays\footnote{The inclusion of charge-conjugated processes is implied throughout this Letter, except where asymmetries are discussed.} across the Dalitz plot~\cite{Dalitz:1953cp,Fabri:1954zz}, which provides a representation of the two-dimensional phase space for the decays, has been previously performed by the BaBar collaboration~\cite{Aubert:2005sk,BaBarpipipi}. A model-independent analysis by the LHCb collaboration, with over an order of magnitude more signal decays and much better signal purity compared to the BaBar data sample, subsequently observed an intriguing pattern of \CP violation in its phase space, notably in regions not associated to any known resonant structure~\cite{LHCb-PAPER-2014-044,LHCb-PAPER-2013-051}.
The observed variation of the \CP asymmetry across the Dalitz plot is expected to be related to the changes in strong phase associated with hadronic resonances, but, to date, has not yet been explicitly described with an amplitude model.
Many phenomenological studies~\cite{Xu:2013rua,Bhattacharya:2013boa,Bhattacharya:2014eca, Krankl:2015fha, Klein:2017xti, Li:2018lbd, CPTconstraint} have provided possible interpretations of the asymmetries.
Particular attention has been devoted to whether large \CP-violation effects could arise from the interference between the broad low-mass spin-$0$ contributions and the spin-$1$ $\rho(770)^0$ resonance~\cite{Dedonder:2010fg,Zhang:2013oqa,Nogueira:2015tsa,Bediaga:2015mia}, 
from mixing between the $\rho(770)^0$ and $\omega(782)$ resonances~\cite{Guo:2000uc, Wang:2015ula, Cheng:2016shb}, or from $\pi\pi \leftrightarrow \kaon\Kbar$ rescattering~\cite{Soni2005,Dedonder:2010fg,Nogueira:2015tsa,Bediaga:2015mia}.
Further experimental studies are needed to clarify which of these sources are connected to the observed \CP\ asymmetries.

In this Letter, results are reported on the amplitude structure of \decay{\Bp}{\pip\pip\pim} decays, obtained by employing decay models that account for \CP violation.
The results are based on a data sample corresponding to $3\invfb$ of $pp$ collisions at centre-of-mass energies of $7$ and $8\tev$, collected with the LHCb detector.
A more detailed description of the analysis is given in a companion paper~\cite{LHCb-PAPER-2019-017}. The LHCb detector is a single-arm forward spectrometer covering the pseudorapidity range $2 < \eta < 5$, described in detail in Refs.~\cite{Alves:2008zz,LHCb-DP-2014-002}

The selection of signal candidates closely follows the procedure used in the model-independent analysis of the same data sample~\cite{LHCb-PAPER-2014-044}, with minor enhancements.
Events containing candidates are selected online by a trigger~\cite{LHCb-DP-2012-004} that includes a hardware and software stage.
The hardware stage requires either energy deposits in the calorimeters associated to signal particles or a trigger caused by other particles in the event. The software triggers require that the signal tracks come from a secondary vertex consistent with the decay of a \bquark hadron.
In the offline selection, two multivariate algorithms are used to separate the \decay{\Bp}{\pip\pip\pim} signal from background formed from random combinations of tracks, and from other \B\ decays with misidentified final state particles, such as \decay{\Bp}{\Kp\pip\pim}.
Candidates that originate from \decay{\Bp}{\Dzb\pip} with subsequent \decay{\Dzb}{\pip\pim} or misidentified $\Kp\pim$ decays are removed with a veto on both $\pip\pim$ invariant mass combinations.

\begin{figure}[tb]
  \centering
  \includegraphics[width=0.49\linewidth]{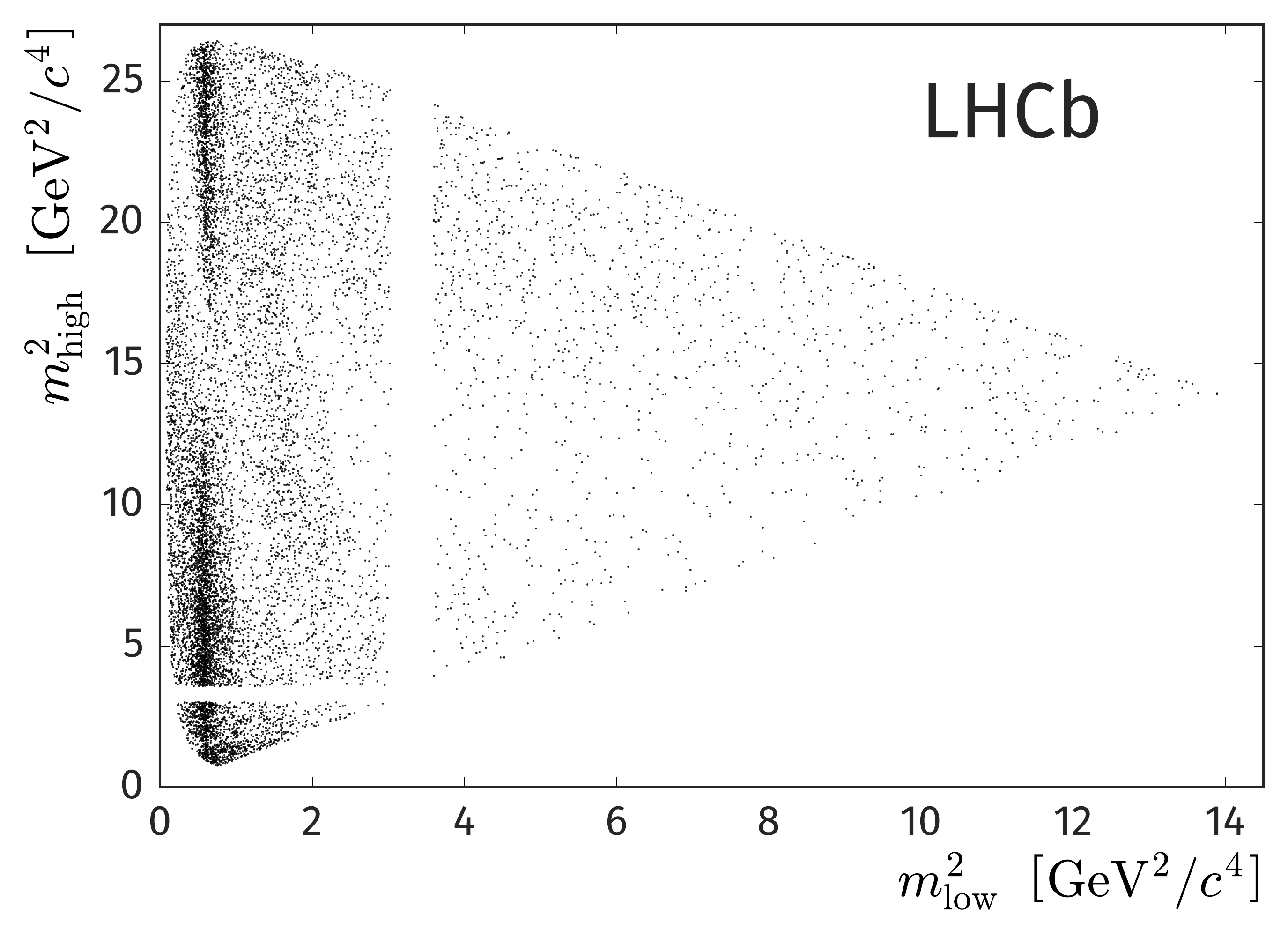}
  \includegraphics[width=0.49\linewidth]{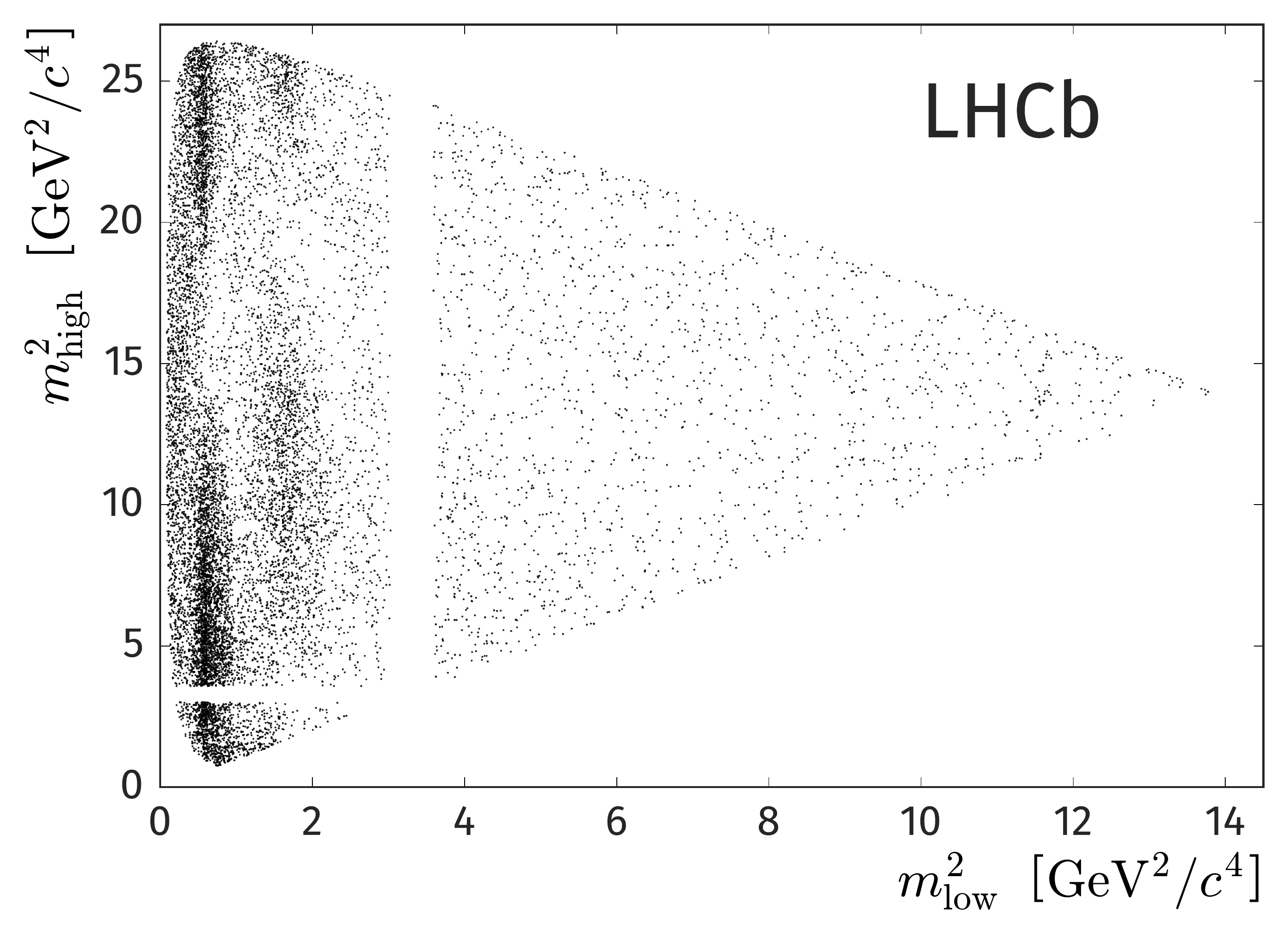}
  \put(-250,140){(a)}
  \put(-25,140){(b)}

  \caption{
    Dalitz-plot distributions for (a)~\Bp and (b)~\Bm candidate decays to $\pipm \pip\pim$. Depleted regions are due to the \Dzb veto.
  }
  \label{fig:dataDP}
\end{figure}

After application of all selection requirements, the $\Bp$-candidate mass distribution is fitted to obtain signal and background yields.
The fit function includes components for signal decays, combinatorial background and misidentified \decay{\Bp}{\Kp\pip\pim} decays.
The signal region in the \Bp candidate mass, $5.249 < m(\pip\pip\pim) < 5.317 \gevcc$, which is used for the Dalitz-plot analysis, is estimated to contain
$20\,600 \pm 1\,600$ signal, $4\,400 \pm 1\,600$ combinatorial background, and $143 \pm 11$ \decay{\Bp}{\Kp\pip\pim} decays,
where the uncertainties reflect the combination of statistical and systematic effects.
The Dalitz-plot distributions of selected \Bp and \Bm candidates are displayed in Fig.~\ref{fig:dataDP}, where the phase space is folded by ordering the $\pip\pim$ pairs by their invariant mass, $m_{\rm low} < m_{\rm high}$.

%% file: Models/models.tex
Given the large number of broad overlapping resonances and decay-channel thresholds, it is particularly challenging to model the \decay{\Bp}{\pip\pip\pim} decay phenomenologically.
Therefore, on top of the conventional ``isobar'' model using a coherent sum of all non-zero spin
resonances, three complementary approaches are used to describe the S-wave amplitude. The first continues in the isobar approach, comprising
the coherent sum of a $\sigma$ pole~\cite{cpole} together with a $\pi\pi \leftrightarrow \kaon\Kbar$ rescattering term~\cite{IgnacioCPT}; the second uses
the K-matrix formalism with parameters obtained from scattering data~\cite{Dalitz:1960du, Aitchison:1972ay, Anisovich:2002ij}; and the third
implements a ``quasi-model-independent'' (QMI) approach, inspired by previous QMI analyses~\cite{Aitala:2005yh}, where the dipion mass spectrum is divided into bins with independent magnitudes and phases that are free to vary in the amplitude fit.

The amplitude for \Bp and \Bm signal decays is constructed as the sum over $N$ resonant contributions and the S-wave component,
\begin{equation}
\label{eq:isobar}
A^{\pm}(m^2_{13}, m^2_{23}) = \sum_{j=1}^{N} c_j^{\pm} F_j(m^2_{13}, m^2_{23}) + A^{\pm}_{\rm{S}}(m^2_{13}, m^2_{23}) \,,
\end{equation}
where $m_{13}$ and $m_{23}$ denote the $\pi^+\pi^-$ invariant mass combinations.
Bose symmetry is accounted for by enforcing the amplitude to be identical under interchange of the two like-sign pions, making the labelling of the two combinations arbitrary.
The $F_j$ term is the normalised dynamical amplitude
of resonance $j$, represented by a mass lineshape multiplied by the spin-dependent angular distribution using the Zemach
tensor formalism~\cite{Zemach:1963bc,Zemach:1968zz} and Blatt--Weisskopf barrier factors~\cite{blatt-weisskopf}.
The complex coefficients, $c_j^{\pm}$, give the relative contribution of each resonance, and $A^{\pm}_{\rm{S}}$ is the S-wave amplitude (isobar, K-matrix or QMI).
The amplitude models account for \CP-violating differences between the distributions of \Bp and \Bm decays by allowing the $c_j^{\pm}$ coefficients, and relevant parameters in $A^{\pm}_{\rm{S}}$, to take different values in the two cases.
A likelihood function is constructed from the squared magnitude of the signal amplitude, accounting for efficiency effects and normalisation, and including background contributions modelled from data sidebands and simulation.
The signal parameters are evaluated in the fit by minimising the negative logarithm of the total likelihood, calculated for all candidates in the signal region.
The \texttt{Laura$^{++}$} package~\cite{Back:2017zqt} is used for the isobar and K-matrix approaches, while a GPU-accelerated version of the \texttt{Mint2} fitter~\cite{jeremysbrain} is used for the QMI approach.

With the exception of the S-wave, the included components are identical in each approach and consist  of the $\rho(770)^0$ and $\omega(782)$ resonances described by a coherent $\rho$--$\omega$ mixing model~\cite{rhoomega},
and the $f_2(1270)$, $\rho(1450)^0$, and $\rho_3(1690)^0$ resonances.
These latter three resonances are all described by relativistic Breit--Wigner lineshapes.
The choice of which resonances to include is made starting from the model obtained in the BaBar analysis~\cite{BaBarpipipi}, with additional contributions included if they cause a significant improvement in
the fit to data.

In each approach, model coefficients for $B^+$ and $B^-$ decays are obtained simultaneously.
The amplitude coefficients extracted from the fit, $c_j^{\pm} = (x \pm \delta x) + i(y \pm \delta y)$, where positive (negative) signs are used for $B^+$ ($B^-$) decays, are defined such that \CP violation is permitted.
For the dominant $\rho$--$\omega$ mixing component, the magnitude of the coefficient in the $B^+$ amplitude is fixed to unity to set the scale, while both $B^+$ and $B^-$ coefficients are aligned to the real axis as the absolute phase carries no physical meaning.

%% file: Results/results.tex
Good overall agreement between the data and the model is obtained for all three S-wave approaches, with some localised discrepancies that are discussed below.
Moreover, the values for the \CP-averaged fit fractions and quasi-two-body \CP asymmetries (rate asymmetries between a quasi-two-body decay and its \CP conjugate), derived from the fit coefficients and given in Table~\ref{tab:sysresults}, show good agreement between the three approaches.

\begin{table*}[tb]
\begin{center}
\caption{\small
Results for \CP-conserving fit fractions, quasi-two-body \CP asymmetries, and phases for each component relative to the combined $\rho(770)^0$--$\omega(782)$ model, given for each S-wave approach. The $\rho(770)^0$ and $\omega(782)$ values are extracted from the combined $\rho(770)^0$--$\omega(782)$ mixing model. The first uncertainty is statistical while the second is systematic.
}
\label{tab:sysresults}
\resizebox{1.0\textwidth}{!}{
\begin{footnotesize}
  \input{tabs/tab_systematics_all.tex}
\end{footnotesize}
}
\end{center}
\end{table*}

\begin{figure}[tb]
  \begin{center}
    \includegraphics[width=0.46\linewidth]{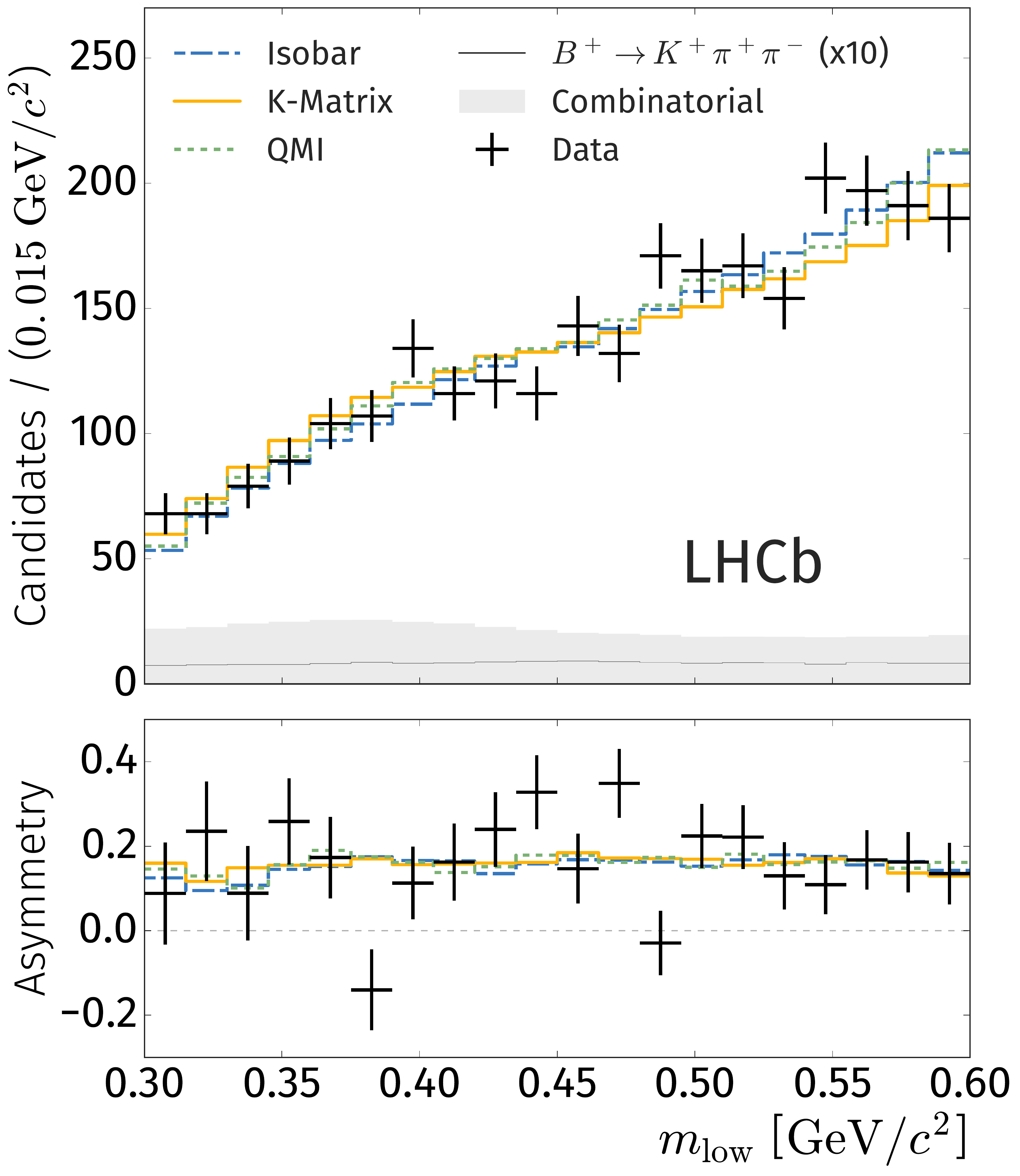}
    \includegraphics[width=0.445\linewidth]{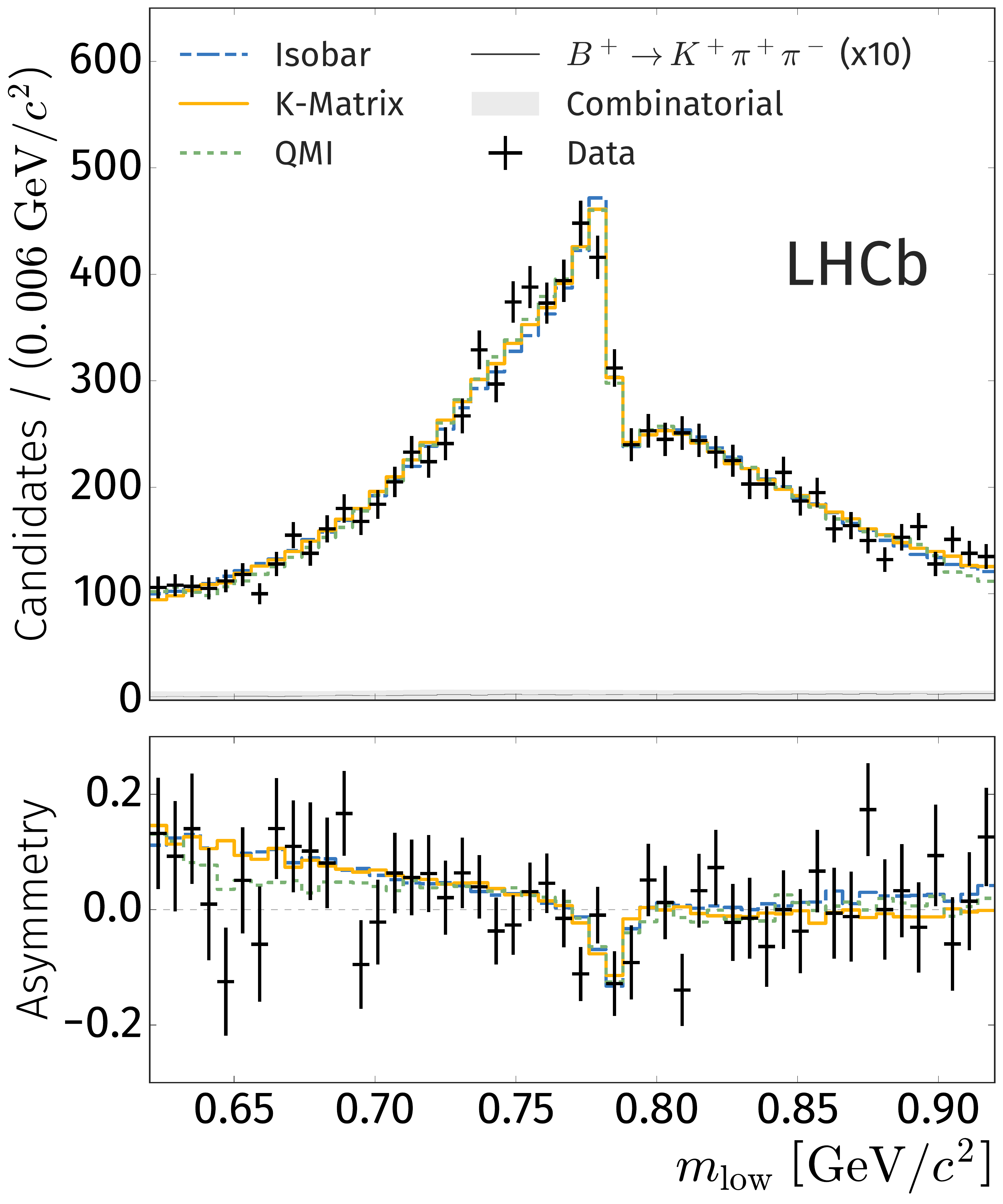}
    \put(-383,30){(a)}
    \put(-170,30){(b)}
  \end{center}
  \caption{
    \small
    Projections of data and fits (top) on $m_{\rm low}$ in (a)~the low $m(\pip\pim)$ region and (b)~the \mbox{$\rho$--$\omega$} region, with (bottom) the corresponding \CP asymmetries in these ranges.
  }
  \label{fig:rhoProj1}
\end{figure}

\begin{figure}[tb]
  \begin{center}
    \includegraphics[width=0.46\linewidth]{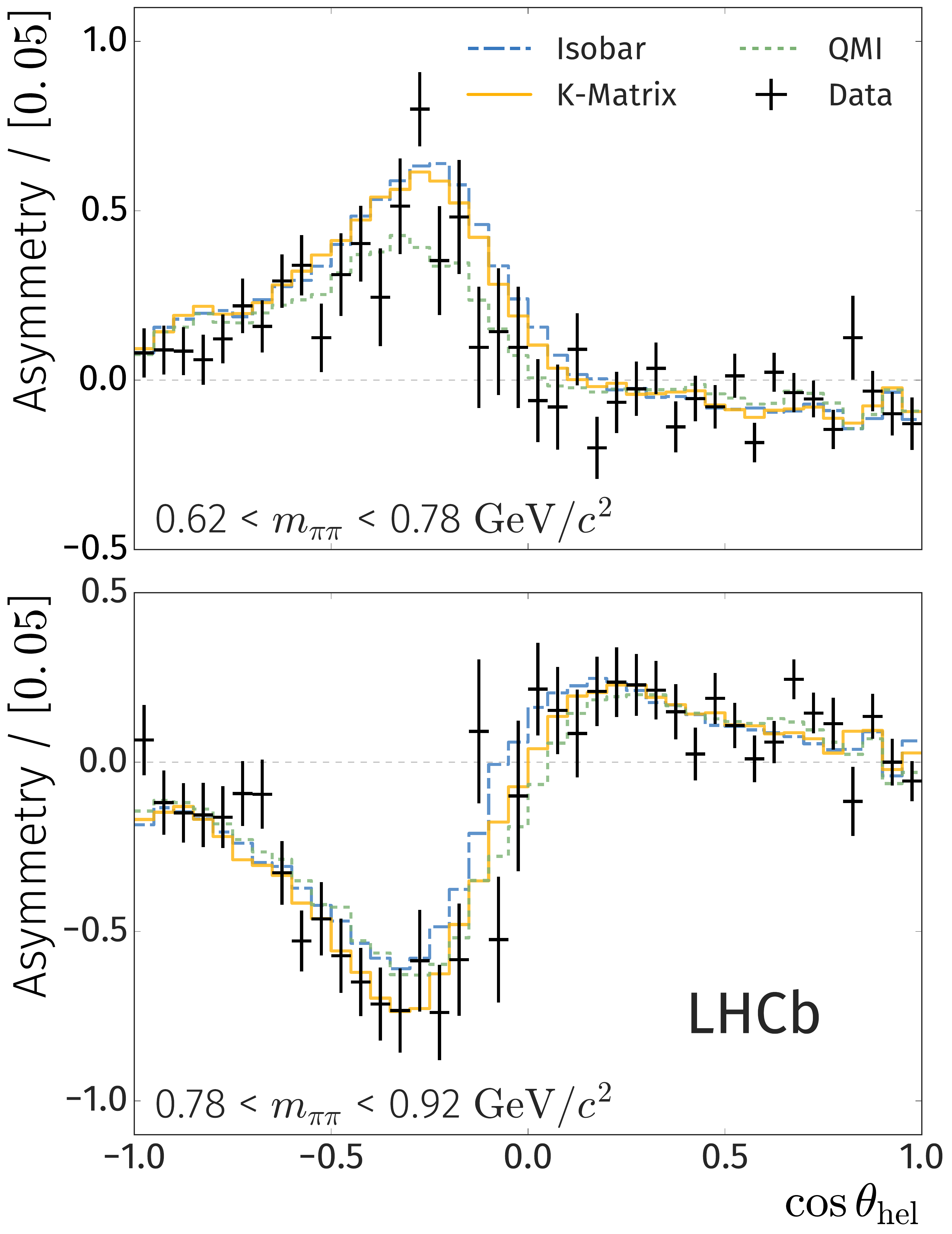}
    \put(-175,167){(a)}
    \put(-175,40){(b)}
  \end{center}
  \caption{
    \small
    Projections of the \CP asymmetry for data and fits as a function of $\cos\theta_{\rm hel}$ in the regions (a)~below and (b)~above the $\rho(770)^0$ resonance pole.
  }
  \label{fig:rhoProj2}
\end{figure}

\begin{figure}[tb]
  \begin{center}
    \includegraphics[width=0.46\linewidth]{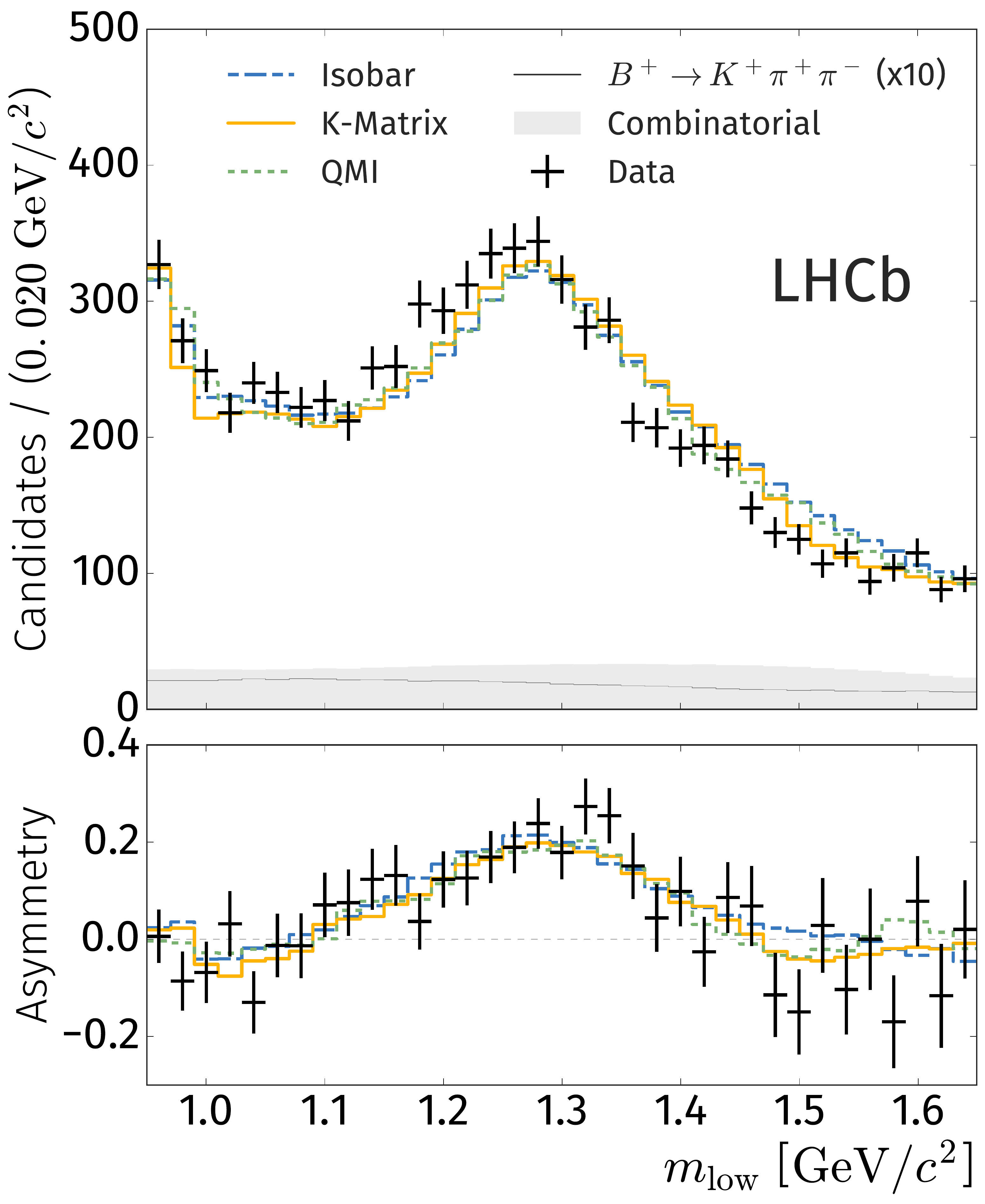}
  \end{center}
  \caption{
   \small
   Projections of data and fits (top) on $m_{\rm low}$ in the $f_2(1270)$ mass region, with (bottom) the corresponding \CP asymmetry.
  }
  \label{fig:massProj}
\end{figure}

Projections of the data and the fit models are shown in regions of the data with $m(\pip\pim) < 1 \gevcc$ in Fig.~\ref{fig:rhoProj1}.
The $\rho(770)^0$ resonance is found to be the dominant component in all models, with a fit fraction of around $55\%$ and a quasi-two-body \CP asymmetry that is consistent with zero.
The effect of $\rho$--$\omega$ mixing is very clear in the data (Fig.~\ref{fig:rhoProj1}(b)) and is well described by the models.
Contrary to some theoretical predictions~\cite{Guo:2000uc, Wang:2015ula, Cheng:2016shb}, there is no evident \CP-violation effect associated with $\rho$--$\omega$ mixing.
However, a clear \CP asymmetry is seen at values of $m(\pip\pim)$ below the $\rho(770)^0$ resonance, where only the S-wave amplitude contributes significantly (Fig.~\ref{fig:rhoProj1}(a)).
A detailed inspection of the behaviour of the S-wave, given in Ref.~\cite{LHCb-PAPER-2019-017}, shows that this \CP asymmetry remains approximately constant up to the inelastic threshold $2m_K$, where it appears to change sign; this is seen in all three approaches to the S-wave description.
Estimates of the significance of this \CP-violation effect give values in excess of ten Gaussian standard deviations ($\sigma$) in all the S-wave models. These estimates are obtained from the change in negative log-likelihood between, for each S-wave approach, the baseline fit and alternative fits where no such \CP violation is allowed.

An additional source of \CP violation, associated principally with the interference between S- and P-waves, is clearly visible when inspecting the $\cos\theta_{\rm hel}$ distributions separately in regions above and below the $\rho(770)^0$ peak (Fig.~\ref{fig:rhoProj2}(a) and~(b)).
Here, $\theta_{\rm hel}$ is the angle, evaluated in the $\pip\pim$ rest frame, between the pion with opposite charge to the \B and the third pion from the \B decay.
These asymmetries are modelled well in all three approaches to the S-wave description.
Evaluation of the significance of \CP violation in the interference between S- and P-waves gives values in excess of $25\sigma$ in all the S-wave models.

At higher $m(\pip\pim)$ values, the $f_2(1270)$ component is found to have a \CP-averaged fit fraction of around $9\%$ and a very large quasi-two-body \CP asymmetry of around $40\%$, as can be seen in Fig.~\ref{fig:massProj} and Table~\ref{tab:sysresults}. This is the first observation of \CP violation in any process involving a tensor resonance. The central value of the \CP asymmetry is consistent with some theoretical predictions~\cite{Li:2018lbd,Cheng:2010yd,Zou:2012td} that, however, have large uncertainties.
The significance of \CP violation in the complex amplitude coefficients of the $f_2(1270)$ component is in excess of $10\sigma$.
This conclusion holds in all the S-wave models and is robust against variations of the models performed to evaluate systematic uncertainties.

The parameters associated to the $\rho(1450)^0$ and $\rho_3(1690)^0$ resonances agree less well, but are nevertheless broadly consistent, between the different models.
The small $\rho_3(1690)^0$ contribution exhibits a large quasi-two-body \CP asymmetry; however this result is subject to significant systematic uncertainties, particularly due to ambiguities in the amplitude model, and therefore is not statistically significant.

The main sources of experimental systematic uncertainty are related to the signal, combinatorial and peaking background parameterisation in the \Bp invariant-mass fit, and the description of the efficiency variation across the Dalitz plot.
Also considered, and found to be numerically larger for most results, are systematic uncertainties related to the physical amplitude models.
These comprise the variation of masses and widths, according to the world averages~\cite{PDG2018}, of established resonances, in addition to the inclusion of more speculative resonant structures.
A small contribution from the $\rho(1700)^0$ resonance is expected by some theory predictions~\cite{Li:2017mao} and is considered a source of systematic uncertainty since the inclusion of this term did not significantly improve the models' agreement with data.

A clear discrepancy between all three modelling approaches and the data can be observed in the $f_2(1270)$ region (Fig.~\ref{fig:massProj}).
This discrepancy can be resolved by freeing the $f_2(1270)$ mass parameter in the fit, however, the values obtained are significantly different from the world-average value.
The discrepancy could arise from interference with an additional spin-2 resonance in this region, but all well established states are either too high in mass or too narrow in width to be likely to cause a significant effect.
The inclusion of a second spin-2 component in this region, with free mass and width parameters, results in values of the $f_2(1270)$ mass consistent with the world average, where parameters of the additional state are broadly consistent with those of the speculative $f_2(1430)$ resonance; however the values obtained for the mass and width of the additional state are inconsistent between fits with different approaches to the S-wave description.
Subsequent analysis of larger data samples will be required to obtain a more detailed understanding of the $\pi\pi$ D-wave in $\Bp \to \pip\pip\pim$ decays.
Variation of the $f_2(1270)$ mass with respect to the world-average value, along with the addition of a second spin-2 resonance in this region, are taken into account in the systematic uncertainties.

%% file: tabs/tab_systematics_all.tex
\renewcommand{\arraystretch}{1.25}
\begin{tabular}
{@{\hspace{0.5cm}}l@{\hspace{0.25cm}}  @{\hspace{0.25cm}}c@{\hspace{0.25cm}}  @{\hspace{0.25cm}}c@{\hspace{0.25cm}}  @{\hspace{0.25cm}}c@{\hspace{0.25cm}}  @{\hspace{0.25cm}}c@{\hspace{0.5cm}}}
      \hline \hline
      Contribution    & Fit fraction ($10^{-2}$)        & $A_{\CP}$ ($10^{-2}$)               & \Bp phase ($^{\circ}$) & \Bm phase ($^{\circ}$) \\ \hline
      Isobar model & \\[3pt]
      \hspace{10pt}$\rho(770)^0$     & $55.5\phantom{0} \pm 0.6\phantom{0} \pm 2.5\phantom{0}$  & $\phantom{0} {+}0.7 \pm \phantom{0}1.1 \pm \phantom{0}1.6$     & --- & ---                       \\
      \hspace{10pt}$\omega(782)$   & $\phantom{0}0.50 \pm 0.03 \pm 0.05$   & $\phantom{0} {-}4.8 \pm \phantom{0}6.5 \pm \phantom{0}3.8$    & $\phantom{0}{-}19 \pm \phantom{0}6 \pm {} \phantom{00}1$ & $\phantom{00}{+}8 \pm \phantom{0}6 \pm {} \phantom{00}1$     \\
      \hspace{10pt}$f_2(1270)$     & $\phantom{0}9.0\phantom{0} \pm 0.3\phantom{0} \pm 1.5\phantom{0}$   & $+46.8 \pm \phantom{0}6.1 \pm \phantom{0}4.7$    & $\phantom{00}{+}5 \pm \phantom{0}3 \pm \phantom{0}12$ & $\phantom{0} {+}53 \pm \phantom{0} 2 \pm \phantom{0}12$     \\
      \hspace{10pt}$\rho(1450)^0$    & $\phantom{0}5.2\phantom{0} \pm 0.3\phantom{0} \pm 1.9\phantom{0}$   & $-12.9 \pm \phantom{0}3.3 \pm 35.9$  & $+127 \pm \phantom{0}4 \pm \phantom{0}21$ & $+154 \pm \phantom{0}4 \pm \phantom{00}6$   \\
      \hspace{10pt}$\rho_3(1690)^0$  & $\phantom{0}0.5\phantom{0} \pm 0.1\phantom{0} \pm 0.3\phantom{0}$   & $-80.1 \pm 11.4 \pm 25.3$ & $\phantom{0}{-}26 \pm \phantom{0}7 \pm \phantom{0}14$ & $\phantom{0}{-}47 \pm 18 \pm \phantom{0}25$ \\
      \hspace{10pt}S-wave          & $25.4\phantom{0} \pm 0.5\phantom{0} \pm 3.6\phantom{0}$  & $+14.4 \pm \phantom{0}1.8 \pm \phantom{0}2.1$    & --- & --- \\[3pt]
      \hspace{20pt}Rescattering         & $\phantom{0}1.4\phantom{0} \pm 0.1\phantom{0} \pm 0.5\phantom{0}$   & $+44.7 \pm \phantom{0}8.6 \pm 17.3$   & $\phantom{0}{-}35 \pm \phantom{0}6 \pm \phantom{0}10$ & $\phantom{00}{-}4 \pm \phantom{0}4 \pm \phantom{0}25$ \\
      \hspace{20pt}$\sigma$        & $25.2\phantom{0} \pm 0.5\phantom{0} \pm 5.0\phantom{0}$  & $+16.0 \pm \phantom{0}1.7 \pm \phantom{0}2.2$    & $+115 \pm \phantom{0}2 \pm \phantom{0}14$ & $+179 \pm \phantom{0}1 \pm \phantom{0}95$ \\

      K-matrix & \\[3pt]

      \hspace{10pt}$\rho(770)^0$     & $56.5\phantom{0} \pm 0.7\phantom{0} \pm 3.4\phantom{0}$  & $\phantom{0} {+}4.2 \pm \phantom{0}1.5 \pm \phantom{0}6.4$     & --- & ---                      \\
      \hspace{10pt}$\omega(782)$   & $\phantom{0}0.47 \pm 0.04 \pm 0.03$   & $\phantom{0} {-}6.2 \pm \phantom{0}8.4 \pm \phantom{0}9.8$    & $\phantom{0}{-}15 \pm \phantom{0}6 \pm \phantom{00}4$ & $\phantom{00}{+}8 \pm \phantom{0}7 \pm \phantom{00}4$     \\
      \hspace{10pt}$f_2(1270)$     & $\phantom{0}9.3\phantom{0} \pm 0.4\phantom{0} \pm 2.5\phantom{0}$   & $ {+}42.8 \pm \phantom{0}4.1 \pm \phantom{0}9.1$    & $\phantom{0}{+}19 \pm \phantom{0}4 \pm \phantom{0}18$ & $\phantom{0}{+}80 \pm \phantom{0}3 \pm \phantom{0}17$     \\
      \hspace{10pt}$\rho(1450)^0$    & $10.5\phantom{0} \pm 0.7\phantom{0} \pm 4.6\phantom{0}$  & $\phantom{0} {+}9.0 \pm \phantom{0}6.0 \pm 47.0$    & ${+}155 \pm \phantom{0}5 \pm \phantom{0}29$ & ${-}166 \pm \phantom{0}4 \pm \phantom{0}51$   \\
      \hspace{10pt}$\rho_3(1690)^0$  & $\phantom{0}1.5\phantom{0} \pm 0.1\phantom{0} \pm 0.4\phantom{0}$   & ${-}35.7 \pm 10.8 \pm 36.9$ & $\phantom{0}{+}19 \pm \phantom{0}8 \pm \phantom{0}34$ & $\phantom{00}{+}5 \pm \phantom{0}8 \pm \phantom{0}46$ \\
      \hspace{10pt}S-wave          & $25.7\phantom{0} \pm 0.6\phantom{0} \pm 3.0\phantom{0}$  & $+15.8 \pm \phantom{0}2.6 \pm \phantom{0}7.2$    & --- & --- \\[3pt]

      QMI & \\[3pt]

      \hspace{10pt}$\rho(770)^0$     & $54.8\phantom{0} \pm 1.0\phantom{0} \pm 2.2\phantom{0}$  & $\phantom{0} {+}4.4 \pm \phantom{0}1.7 \pm \phantom{0}2.8$     & --- & ---                      \\
      \hspace{10pt}$\omega(782)$   & $\phantom{0}0.57 \pm 0.10 \pm 0.17$   & $\phantom{0} {-}7.9 \pm 16.5 \pm 15.8$  & $\phantom{0}{-}25 \pm \phantom{0}6 \pm \phantom{0}27$ & $\phantom{00}{-}2 \pm \phantom{0}7 \pm \phantom{0}11$     \\
      \hspace{10pt}$f_2(1270)$     & $\phantom{0}9.6\phantom{0} \pm 0.4\phantom{0} \pm 4.0\phantom{0}$   & $+37.6 \pm \phantom{0}4.4 \pm \phantom{0}8.0$    & $\phantom{0}{+}13 \pm \phantom{0}5 \pm \phantom{0}21$ & $\phantom{0}{+}68 \pm \phantom{0}3 \pm \phantom{0}66$     \\
      \hspace{10pt}$\rho(1450)^0$    & $\phantom{0}7.4\phantom{0} \pm 0.5\phantom{0} \pm 4.0\phantom{0}$   & $-15.5 \pm \phantom{0}7.3 \pm 35.2$   & $+147 \pm \phantom{0}7 \pm 152$ & $ -175 \pm \phantom{0}5 \pm 171$   \\
      \hspace{10pt}$\rho_3(1690)^0$  & $\phantom{0}1.0\phantom{0} \pm 0.1\phantom{0} \pm 0.5\phantom{0}$   & $-93.2 \pm \phantom{0}6.8 \pm 38.9$  & $\phantom{00}{+}8 \pm 10 \pm \phantom{0}24$ & $\phantom{0}{+}36 \pm 26 \pm \phantom{0}46$ \\
      \hspace{10pt}S-wave          & $26.8\phantom{0} \pm 0.7\phantom{0} \pm 2.2\phantom{0}$  & $+15.0 \pm \phantom{0}2.7 \pm \phantom{0}8.1$    & --- & ---\\
      \hline \hline
\end{tabular}

%% file: Conclusions/conclusions.tex
In summary, an amplitude analysis of the ${\decay{\Bp}{\pip\pip\pim}}$ decay is performed with data corresponding to $3$\invfb of LHCb Run 1 data, using three complementary approaches to describe the large S-wave contribution to this decay.
Good agreement is found between all three models and the data.
In all cases, significant \CP violation is observed in the decay amplitudes associated with the $f_2(1270)$ resonance and with the $\pip\pim$ S-wave at low invariant mass, in addition to \CP violation characteristic of interference between the spin-$1$ $\rho(770)^0$ resonance and the spin-$0$ S-wave contribution.
Violation of \CP symmetry is previously unobserved in these processes and, in particular, this is the first observation of \CP violation in the interference between two quasi-two-body decays.
As such, these results provide significant new insight into how \CP violation manifests in multi-body \B-hadron decays, and motivate further study into the processes that govern \CP violation at low $\pi\pi$ invariant mass.

%% file: acknowledgements.tex
\section*{Acknowledgements}
%
% These Acknowledgements valid from 3-May-2019
%
\noindent We express our gratitude to our colleagues in the CERN
accelerator departments for the excellent performance of the LHC. We
thank the technical and administrative staff at the LHCb
institutes.
We acknowledge support from CERN and from the national agencies:
CAPES, CNPq, FAPERJ and FINEP (Brazil); 
MOST and NSFC (China); 
CNRS/IN2P3 (France); 
BMBF, DFG and MPG (Germany); 
INFN (Italy); 
NWO (Netherlands); 
MNiSW and NCN (Poland); 
MEN/IFA (Romania); 
MSHE (Russia); 
MinECo (Spain); 
SNSF and SER (Switzerland); 
NASU (Ukraine); 
STFC (United Kingdom); 
DOE NP and NSF (USA).
We acknowledge the computing resources that are provided by CERN, IN2P3
(France), KIT and DESY (Germany), INFN (Italy), SURF (Netherlands),
PIC (Spain), GridPP (United Kingdom), RRCKI and Yandex
LLC (Russia), CSCS (Switzerland), IFIN-HH (Romania), CBPF (Brazil),
PL-GRID (Poland) and OSC (USA).
We are indebted to the communities behind the multiple open-source
software packages on which we depend.
Individual groups or members have received support from
AvH Foundation (Germany);
EPLANET, Marie Sk\l{}odowska-Curie Actions and ERC (European Union);
ANR, Labex P2IO and OCEVU, and R\'{e}gion Auvergne-Rh\^{o}ne-Alpes (France);
Key Research Program of Frontier Sciences of CAS, CAS PIFI, and the Thousand Talents Program (China);
RFBR, RSF and Yandex LLC (Russia);
GVA, XuntaGal and GENCAT (Spain);
the Royal Society
and the Leverhulme Trust (United Kingdom).

%% file: LHCb_Authorship_12-Mar-2019.tex
% LHCb collaboration author list
% Data extracted on July 29th, 2019 at 10:36am for reference date 12-Mar-2019
\centerline
{\large\bf LHCb collaboration}
\begin
{flushleft}
\small
R.~Aaij$^{30}$,
C.~Abell{\'a}n~Beteta$^{47}$,
B.~Adeva$^{44}$,
M.~Adinolfi$^{51}$,
C.A.~Aidala$^{78}$,
Z.~Ajaltouni$^{8}$,
S.~Akar$^{62}$,
P.~Albicocco$^{21}$,
J.~Albrecht$^{13}$,
F.~Alessio$^{45}$
% ...

{\footnotesize \it

% ...
$ ^{8}$Universit{\'e} Clermont Auvergne, CNRS/IN2P3, LPC, Clermont-Ferrand, France\\
% ...
$ ^{13}$Fakult{\"a}t Physik, Technische Universit{\"a}t Dortmund, Dortmund, Germany\\
% ...
$ ^{21}$INFN Laboratori Nazionali di Frascati, Frascati, Italy\\
% ...
$ ^{30}$Nikhef National Institute for Subatomic Physics, Amsterdam, Netherlands\\
% ...
$ ^{44}$Instituto Galego de F{\'\i}sica de Altas Enerx{\'\i}as (IGFAE), Universidade de Santiago de Compostela, Santiago de Compostela, Spain\\
$ ^{45}$European Organization for Nuclear Research (CERN), Geneva, Switzerland\\
% ...
$ ^{47}$Physik-Institut, Universit{\"a}t Z{\"u}rich, Z{\"u}rich, Switzerland\\
% ...
$ ^{51}$H.H. Wills Physics Laboratory, University of Bristol, Bristol, United Kingdom\\
% ...
$ ^{62}$University of Cincinnati, Cincinnati, OH, United States\\
% ...
$ ^{65}$Syracuse University, Syracuse, NY, United States\\
% ...
$ ^{78}$University of Michigan, Ann Arbor, United States, associated to $^{65}$\\
\bigskip
% ...
}
\end{flushleft}